\documentstyle[12pt]{article}
\begin{document}
\baselineskip 22pt plus2pt

\noindent \hspace*{9cm}TECHNION-PH-93-13\\
\hspace*{9cm} hep-th/9304059\\
\hspace*{9cm} April 1993
\vspace{1cm}

\begin{center}
{\bf DISSIPATION OF THE STRING EMBEDDING DIMENSION IN THE SINGLET SECTOR
OF A MATRIX MODEL AT LARGE $\alpha^\prime$}
\vspace{0.50cm}

Joshua Feinberg\footnote{Supported in part by the Henri Gutwirth
Fund}\footnote{Bitnet: PHR74JF@Technion.bitnet}
\end{center}
\vspace{0.50cm}

\begin{center}
{\bf Abstract}
\end{center}
\vspace{0.25cm}

\noindent\parbox{14cm}
{The one dimensional Fermi gas of matrix eigenvalues of the Marinari-Parisi
model at positive values of the cosmological constant is generalised.
The number of matrix eigenvalues (i.e. gas particles) is varied while
keeping the effective potential fixed. This model exhibits a transition
from a phase whose continuum behaviour is that of $c=1$ conformal matter
coupled to gravity to the well known pure gravity phase of the original model.
The former phase is characterised by an extremely large Regge slope
$\alpha^\prime$ which scales as $\beta^{2/5}$ causing the scaling regions of
the two phases to overlap. In this way a continuous flow from one phase to
the other is made possible. This phase transition occurs in the
{\em singlet sector} of the matrix model. The density of states and the two
puncture correlator at non zero momenta are calculated on the sphere and
are found to behave very differently in the two phases,a fact which
demonstrates the phase transition. We comment on a possible relation between
this transition and large $\alpha^\prime$ semiclassical expansions in the
continuum.}\\

\break

\noindent{\bf I. \ Introduction}\\

Pure two dimensional quantum gravity is described in the stochastically
stabilised model\cite{GH} by extracting the double scaling (ds)
behaviour\cite{DSL,DSLREV} of the ground state of a forward
Fokker-Planck hamiltonian ${\cal H}$ \cite{MP,KM,AMB} which generates
matrix evolution in (the Euclidean) Langevin time.  This definition of
pure gravity $(k=2)$ may be generalised to all $(2,2k-1)$ multicritical
points\cite{JF1,JF2,MIRAM} providing a non-perturbative consistent
formulation of these models, compatible with the genus expansion.  The
ground state of ${\cal H}$ is clearly a $U(N)$ singlet.  The $U(N)$
singlet sector of ${\cal H}$ describes a one dimensional
Fermi gas\cite{BIPZ} containing $N$ particles.  In the case of pure
gravity (represented by a cubic zero dimensional matrix potential) the
particles are noninteracting and move in an external effective potential
$V_N$ which depends explicitly on $N$\cite{MP,KM}\footnote{In case of
higher multicriticalities $k \geq 3$, the particles are subject to long
range two body interactions\cite{JF1,JF2,MIRAM}}.  For positive values
of the cosmological constant $V_N$ has a pair of adjacent local
quadratic minimum and maximum points (see Fig.~(1)) whose separation
and height difference scale as $N^{-2/5}$ and $N^{-6/5}$, respectively.

The Fermi energy $\epsilon_F$ is not a free parameter and must be fixed
selfconsistently by using the explicit $N$ dependence of $V_N$.
$\epsilon_F$ turns out to coincide with the local minimum of $V_N$ to
leading order in $1/N$\cite{MP,KM}, ensuring a one
cut solution of the planar macroscopic loop\cite{FDAV,JF1,JF2}.  This
leads to a supersymmetric invariant ground state at the planar level
(which is actually broken by non perturbative effects).

The theory of $c=1$ matter coupled to gravity, which is equivalent to a
noncritical string theory propagating in two dimensions, is the
richest
exactly solvable theory of its sort.  As in the previous case, it is
described by the large-$N$ limit of an hermitean matrix (Euclidean)
quantum mechanics\cite{C1,C1REV}.  The singlet sector of the latter is
equivalent to a one dimensional Fermi gas of non interacting particles.
Here, however, the Fermi energy is an arbitrary parameter, and
criticality is attained when the latter reaches a quadratic maximum of
the matrix potential.   In the uncompactified $c=1$ model, as opposed to
the $c<1$ case, there exist infinitely many primary fields labelled by
continuous momenta.  The scaling parameter near the $c=1$ critical point
$\mu$ is related to the cosmological constant $t$ via $t \sim
\mu\ln\mu$.  Correlators in the matrix model scale in $\mu$ with
continuous momentum dependent scaling dimensions.  In particular, the
two point function behaves as
$F(p)\mu^{\sqrt{\alpha^\prime}\mid p\mid}$ \cite{TWOPOINT1} where $F(p)$
is a form factor which contains evenly spaced poles
\cite{TWOPOINT2}corresponding to discrete states\cite{SPECIAL} at real
values of $p$.

Since both matrix models considered above are defined in the same
fashion, it is interesting to investigate the possibility of a flow from
one to the other in the same matrix quantum mechanical model.

A transition from the $c=1$ phase to the pure gravity one, which is of
the Berezinskii-Kosterlitz-Thouless (BKT) type\cite{BKT}, is well known
in the framework of quantum mechanical matrix model compactified along a
circle of radius $R$\cite{FLOW,C1REV}.  Due to compactification, vortex
degrees of freedom (which are associated with {\em non singlet} sectors
of the matrix hamiltonian) exist in the model.  These vortices become
relevant and condensate as the compactification radius is reduced below
a critical value $R_c=2\sqrt{\alpha^\prime}$, corresponding to the high
temperature phase of the model.  This condensation disorders the
embedding dimension and the system becomes zero dimensional.  For
$R>R_c$ on the other hand, the compactified theory is essentially the
same as the uncompactified one.  Equivalently, in the dual picture to
the
circle compactification, the string is embedded into a one dimensional
spatial lattice, whose spacing $\epsilon$ is
$2\pi\alpha^\prime/R$\cite{DISCR,C1REV,FLOW}.    The critical lattice
spacing is $\epsilon_c=\pi\sqrt{\alpha^\prime}$.  When $\epsilon <
\epsilon_c$ (corresponding to $R > R_c$ in the dual picture) the
discretised theory is equivalent to the continuum formulation.  However,
for $\epsilon > \epsilon_c$, lattice points disentangle dynamically and
the matrix model partition function factors into a product of
independent ordinary matrix integrals, describing the pure gravity
phase\footnote{A general discussion of discretised quantum mechanics in
the operatorial formulation, which uses the method of finite elements,
may be found in\cite{FINITE}.  The discretised harmonic oscillator is
treated as a special example, and the effect of dimensional reduction
at gross discretisations is demonstrated in general.}.It is important to
note that these vortices are associated with non singlet sectors of the
matrix model.In fact,the singlet sector partition function of the compactified
model is manifestly dual under inversion of the compactification radius
\cite{FLOW} and is thus free of vortices.

In this paper we generalise the Marinari-Parisi model described above by
embedding it into a grand canonical ensemble -- namely, we introduce an
arbitrary chemical potential $\mu_F$, allowing ${\cal N}$, the number of
particles in the gas, to vary, while maintaining $N$ and $V_N$ fixed.
When ${\cal N}$ matches $N$, as $\mu_F$ and $\epsilon_F$ coincide -- we
return to the pure gravity phase, described by the Marinari-Parisi
model.  Increasing $\mu_F$ above $\epsilon_F$, we eventually reach the
quadratic  local maximum of $V_N$, whose double scaled vicinity
corresponds to the $c=1$ phase, albeit with extremely large Regge slope
$\alpha^\prime$, which diverges as $\beta^{2/5}\sim N^{2/5}$.  This fact
causes the scaling regions of the two phases to overlap.

Varying $\mu_F$ we thus flow freely from one phase to the other.Thus
$\mu_F$ serves as a scaling field whose variation may change the embedding
dimension of our random surfaces. Note,however, that this flow involves
only the {\em singlet} sector of the {\em uncompactified} matrix model, in
contrast to the BKT transition discussed above.

The paper is organised as follows: \ In section (II) we recall aspects
of the Marinari-Parisi model relevant for our discussion and define its
generalisation mentioned above. The crucial role played by $\alpha^\prime$
scaling as $\beta^{2/5}$ is discussed.

In section (III) we calculate the density of states to leading order in
$1/\beta$, i.e., on the sphere, and investigate its behaviour as a
function of $\mu_F$ and $\alpha^\prime$.  Section (IV) is devoted to
evaluation of the two puncture correlator $G^{(2)}(p)$ to leading order
in $1/\beta$ along the lines of \cite{TWOPOINT1,TWOPOINT2}.
$G^{(2)}(p)$ is shown explicitly to flow from one phase to the other, as
$\mu_F$ is varied appropriately.
We discuss our results and draw our conclusions in
section (V).\\

\break

\noindent{\bf II. \ The Generalised Stochastic Matrix Model}\\

The singlet sector of the stochastic matrix hamiltonian describing pure
gravity is equivalent to a one dimensional Fermi gas of $N$
noninteracting particles.  Following \cite{MP,KM} we write this
hamiltonian as \cite{JF1,JF2}
\begin{eqnarray}
{\cal H} = \sum^N_{i=1} h(x_i,p_i) = \sum^N_{i=1}
\left[-\frac{1}{2\beta^2} \frac{\partial^2}{\partial x_i^2} +    
V_N(x_i)\right]
\end{eqnarray}
where the single body external potential is
\begin{eqnarray}
V_N(x) =
\frac{1}{8}(-\frac{1}{2}x^2+x+1)^2+\frac{1}{4}\frac{N}{\beta}(x-1) \ .
\end{eqnarray}
Here $x_i \ \  (1 \leq i \leq N)$ are eigenvalues of the $N \times N$
hermitean matrix $\Phi$, in terms of which the original unstable matrix
integral defining pure gravity reads
\begin{eqnarray}
{\cal Z} (\beta,N) = \int d^{N^2} \Phi \exp \left[-\beta
Tr(-\frac{1}{6}\Phi^3+\frac{1}{2}\Phi^2+\Phi)\right]         
\end{eqnarray}
where $\beta \rightarrow \infty$.

$V_N(x)$ is depicted schematically on Fig.~(1).  Since $V_N(x)$
confines
exactly $N$ noninteracting Fermions, the  Fermi energy $\epsilon_F$
turns out to coincide with the local minimum of $V_N(x)$ \cite{MP,KM} at
$x_0 \approx 2 + \left[\frac{2}{3}(1-\frac{N}{\beta})
\right]^{\frac{1}{2}}$, and the classical turning points at $\epsilon_F$
are $a=-2$ and $b\approx
2-2\left[\frac{2}{3}(1-\frac{N}{\beta})\right]^{\frac{1}{2}}$
\cite{JF1,JF2}. Note also that $V_N(x)$ has a local quadratic maximum at
$x_c\approx2-\left[\frac{2}{3}(1-\frac{N}{\beta})\right]^{\frac{1}{2}}$
in the vicinity of $x_0$\footnote{$x_0,x_c$ and $b$ are given here up to
terms of ${\cal O}(1-\frac{N}{\beta})$.}.

The semiclassical ground state energy of the gas vanishes, as a result
of supersymmetry at the planar level\footnote{Supersymmetry is broken,
of course, at the nonperturbative level, since the zero dimensional
matrix potential in Eq.\ (3) is a cubic polynomial.}.

The single particle hamiltonian $h$ in Eq.\ (1) double scales in the
expected pure gravity manner\cite{DSL} under the
substitutions\cite{MP,KM,AMB}
\begin{eqnarray}
x = 2 + y\beta^{-2/5}, \ \frac{N}{\beta}=1-t\beta^{-4/5} 
\end{eqnarray}
where $t$ is the renormalised cosmological constant.  We will consider
only the case of $t>0$, under which $V_N(x)$ behaves as we have
described above.

Eq.~(4) implies the expansion of $V_N(x)$ around $x=2$, which is the
critical end point of the planar distribution of eigenvalues \cite{JF1,JF2}.

Inserting Eq.~(4) into Eqs.~(1) and (2), the scaled hamiltonian reads
\begin{eqnarray}
h(y,p)=\beta^{-6/5}\left[-\frac{1}{2}\frac{\partial^2}{\partial y^2} +
v(y)\right] + E_0 \equiv \beta^{-6/5}h_{\rm d.s.} +E_0 
\end{eqnarray}
where $E_0 = \frac{3}{8} + {\cal O}(\beta^{-4/5})$, $h_{d.s.}$ is the
double scaled hamiltonian and
\begin{eqnarray}
v(y) = \frac{1}{8}(y^3-2ty)+{\cal O}(\beta^{-2/5}) \ . 
\end{eqnarray}
$v(y)$ is depicted schematically on Fig.~(2).  The local extrema of
$V_N(x)$ at $x_0$ and $x_c$ correspond to the minimum and maximum of
$v(y)$ at $+y_0$ and $-y_0$, respectively, where
$y_0 = (2t/3)^{1/2}$.  The scaled Fermi energy $e_c$ coincides with the
minimum $v(y_0)$ and is found to be
\begin{eqnarray}
e_c=v(y_0) = -\frac{1}{4}(2t/3)^{3/2} < 0 \ .
\end{eqnarray}
The scaled classical turning points at that energy are $y_a =
-4\beta^{2/5} \rightarrow - \infty$ and $y_b = -2y_0$.  Note further
that the global minimum of $V_N(x)$ corresponds to $e\sim-\beta^{6/5}
\rightarrow -\infty$.

Despite the fact that $v(y)$ in Eq.~(6) is bottomless, $h(y,p)$ may be
extended into a one parameter family of self adjoint operators, where
for each choice of the parameter, the spectrum of $h(y,p)$ is purely
discrete and ranges from $-\infty$ to $+\infty$ \cite{AMB2}.  The reason
for this being the fact that the ``classical time of flight'' from any
finite turning point $y(e)$ to $-\infty$ is finite, due to the fast
decrease of $v(y)$ as $y\rightarrow-\infty$.  This makes the spectral
analysis of Eq.~(5) similar in many respects to that of a
Schr\"{o}dinger
hamiltonian inside a rigid finite box.  Thus $y = -\infty$ may be
considered as an ordinary turning point.

We now generalise the Marinari-Parisi model, as we have described in the
introduction. We introduce a chemical potential $\mu_F$ as
a new free parameter which allows us to vary the number of particles
${\cal N}$ in the potential well, holding $N$ and $V_N(x)$ fixed.
Increasing $\mu_F$ above $\epsilon_F$ we add more particles into the
potential well, such that ${\cal N} \geq N$.  Clearly, as $\mu_F$
matches $\epsilon_F$ exactly we return to the pure gravity phase.  On
the other hand, as $\mu_F$ reaches $V_N(x_c)$ closely enough we arrive
at the phase where the matrix model describes $c=1$ matter coupled to
gravity.  Note that the pure gravity phase corresponds to a {\em
specific} value of $\mu_F$, while the $c=1$ phase -- to a continuous set
of $\mu_F$ values.

In order to describe the $c=1$ phase of $V_N(x)$, we have to expand and
double scale ${\cal H}$ around the quadratic maximum $V_N(x_c)$.
Doing so we find\footnote{$V^{\prime\prime}_N(x_c)$ and
$V^{\prime\prime\prime}_N(x_c)$  are given
up to a factor of $1 + {\cal O}(\beta^{-2/5}$)}
\begin{eqnarray}
V_N(x) \approx V_N(x_c) - \frac{3}{8}y_0\beta^{-2/5}(x-x_c)^2 +
\frac{1}{8}(x-x_c)^3+\frac{1}{32}(x-x_c)^4  
\end{eqnarray}

The Regge slope of the string theory described by this matrix
model is inversely proportional to $-V^{\prime\prime}(x_c)$
\cite{C1,C1REV} and we find
\begin{eqnarray}
\alpha^\prime\sim-1/V^{\prime\prime}(x_c) =
4\beta^{2/5}/3y_0 =
4\cdot\left[6(1-\frac{N}{\beta})\right]^{-\frac{1}{2}}   
\end{eqnarray}
where a use of Eq.~(4) has been made.  Thus, $\alpha^\prime$ turns out
to be extremely large and diverges as $\beta^{2/5}$ for {\em finite}
values of the pure gravity cosmological constant $t$.  Equivalently, the
string tension $T = 1/2 \pi\alpha^\prime$ vanishes then as
$\beta^{-2/5}$.

The drastic effect of Eq.~(9) on the space-time interpretation of the
$c=1$ matrix model as a two dimensional critical string can be
demonstrated from the following simple scaling argument: Replacing the
coefficient of the quadratic term in Eq.~(8) by a general factor
$\frac{1}{2\alpha^\prime}$ and defining
\begin{eqnarray}
\mu = V_N(x_c)-\mu_F    
\end{eqnarray}
the one particle hamiltonian $h$ in
Eq.~(1) becomes
\begin{eqnarray}
\mu_F - h = \left[ - \frac{1}{2\beta^2} \frac{\partial^2}{\partial z^2}
+ \frac{1}{2\alpha^\prime} z^2 -\mu\right] -
\frac{1}{8}z^3-\frac{1}{32}z^4                   
\end{eqnarray}
The terms square brackets in Eq.~(11) are the usual hamiltonian used to
describe $c=1$ theories coupled to gravity.  Thus, from the obvious
demand that all terms in these brackets scale alike\cite{C1,C1REV} we
obtain the ordinary scaling behaviour of $c=1$, keeping
$-\alpha^\prime=\frac{1}{\omega^2}$ explicitly as a free parameter:
\begin{eqnarray}
z^2\sim\frac{\sqrt{\alpha^\prime}}{\beta}\sim\alpha^\prime\mu ~~
{\rm hence} ~~ h\sim\mu\sim\frac{1}{\beta\sqrt{\alpha^\prime}}
\end{eqnarray}
implying the $c=1$ string coupling\footnote{Eq.~(12) could be deduced
similarly from keeping $\omega=\frac{i}{\sqrt{\alpha^\prime}}$
explicitly in the genus expansion for the density of
states\cite{C1,C1REV} }
\begin{eqnarray}
g^{(1)}_{st} = \frac{1}{\beta\mu\sqrt{\alpha^\prime}} = {\rm const.} \
\frac{t^{1/4}}{\beta^{6/5}\mu} \ .                    
\end{eqnarray}
For finite values of $\alpha^\prime$ we obtain the usual $c=1$ results.
However, substituting $\alpha^\prime$ from Eq.~(9) scaling properties
change drastically, namely: $z \sim\beta^{-2/5}$,
$h\sim\mu\sim\beta^{-6/5}$ and the cubic term in Eq.~(11) becomes
relevant.  These are precisely the scaling laws of
Eqs.~(4)-(6)\cite{MP,KM,AMB} which correspond to pure gravity.  Indeed,
under the substitution of Eq.~(9), Eq.~(11) is nothing but Eq.~(5) where
the polynomial $v(y)$ in Eq.~(6) is expanded around its maximum
$v(-y_0)$.

Therefore, the scaling regions of one dimensional Fermi gas systems,
describing $c=1$ string and pure gravity, which are infinitely separated
for finite values of $\alpha^\prime$, coincide as $\alpha^\prime$
diverges as in Eq.~(9).
This implies that one can flow from one theory to the other by changing
$\mu$ (or $g^{(1)}_{st}$) in the domain $0 \leq \mu\leq
V_N(x_c)-V_N(x_0)\sim\beta^{-6/5}$.  Thus we expect that the notion
of two dimensional
target space-time of the $c=1$ string will be completely lost for
finite $g^{(1)}_{st}$,as $\alpha^\prime$ diverges.

For a given value of $\mu$,
 $g^{(1)}_{st}$
may grow by increasing $t$.  As $t$ increases, the separation of the two
local extrema of $v(y)$ at $\pm y_0$ increases as
$y_0^{3}\sim t^{3/2}$ (Eq.~(7)).  Clearly, from Eq.~(4) $t$ can
diverge as $\beta^{4/5}$ at most.  In such a case, corresponding to the
perturbative region of pure gravity, the extrema of $v(y)$ become
infinitely distant and the two phases are separated accordingly
\footnote{Equivalently,$V_N(x_c)-V_N(x_0)$ becomes ${\cal O}(1)$}.
Indeed, $\alpha^\prime$ becomes finite in this limit and we recover the
ordinary $c=1$ scaling behaviour.

The flow that we have described above (and will substantiate in the following
sections) occurs solely in the $U(N)$ {\em singlet}
sector of the {\em uncompactified} matrix hamiltonian ${\cal H}$, contrary
to the BKT transition mentioned in the introduction.  Recall, however, that
the critical radius of that transition is $R_c\sim 2\sqrt{\alpha^\prime}$
which diverges as $\beta^{1/5}$ for $\alpha^\prime$ in Eq.~(9).  Thus,
in a circle compactification of our model, nonsinglet $U(N)$ sectors
(and therefore vortices) are relevant for {\em any} radius as
$\beta\rightarrow\infty$\cite{FLOW,C1REV}, i.e., the compactified theory
has only the disordered phase.  Alternatively,the energy gap from the
singlet ground state of the matrix model to the lowest members of the
adjoint multiplet of ${\cal H}$ scales as \cite{FLOW}
$\frac{-1}{2\pi\beta\sqrt{\alpha^\prime}}\ln (\mu\sqrt{\alpha^\prime})$
i.e. as $\frac{\beta^{-6/5}}{2\pi}\ln\beta$ for finite $g^{(1)}_{st}$.
The non singlet states and their associated vortices become relevant for
{\em any} non zero temperature,as low as we like. We may understand this
intuitively from the fact that the string is extremely flexible in our case
(for finite $t$) since the string tension is
$T=1/2\pi\alpha^\prime\sim\beta^{-2/5}$.  It thus winds around any circle
in target space with almost no energy cost.It is thus important to check
whether the singlet sector partition function of our model remains dual under
inversion of the compactification radius,which is the case for the
compactified matrix model at finite values of ${\alpha^\prime}$.This
question will not be discussed here.

At this point it is interesting to note a qualitative analogy of our
model to the finite temperature quantum mechanics of the unharmonic
oscillator discussed in \cite{KURZ} that involves vortex excitations .The
latter,however,do not lead to a BKT transition.

We close this section with the following remark:Note that when $\mu_F$
is varied between the two local extrema of $V_N(x)$,
tunnelling effects may partly fill the small well right to the large
abyss of $V_N(x)$ (Fig.~(1)).These tunnelling effects are nonperturbative
in the string couplings of either phases.  They result in two cut
non-perturbative loop expectations which might be interpretted as leading
in the double scaling limit (Eq.~(5)) to nonvanishing expectations for
macroscopic loops of negative lengths (albeit, decreasing exponentially as
its absolute value is increased)\cite{FDAV}.
But since all these effects are unseen
in either genus expansions we ignore them in this paper.\\

\break

\noindent{\bf III. \ The Density of States on the Sphere}\\

The discussion in the previous section is demonstrated clearly in the
behaviour of the density of states of the gas $\rho(\mu_F)$, computed to
leading order in $1/\beta$, i.e., on the sphere.  We express
$\rho(\mu_F)$
as a function of the classical turning points in the potential $V_N(x)$
in closed form.  Using Eqs.~(4) and (5) the scaled chemical potential
$e$ is defined from\footnote{Since the scaling regions of the $c=1$
string and pure gravity match, we may define the scaled chemical
potential either relative to $V_N(x_c)$ or relative to
$\epsilon_F=V_N(x_0)$.  We choose here the latter option. }
\begin{eqnarray}
\mu_F - \epsilon_F = \mu_F - V_N(x_0) = \beta^{-6/5}e      
\end{eqnarray}
such that
\begin{eqnarray}
\mu_F - V_N(x)=\beta^{-6/5}[e-v(y)]+{\cal O}(\beta^{-8/5})   
\end{eqnarray}

For $e_c<e<-e_c$ where $e_c$ is given in Eq.~(7), we find three real
turning points as the real roots of the cubic
\begin{eqnarray}
\frac{1}{8}(y^3-2ty)-e = 0  \ .     
\end{eqnarray}
The fourth turning point at $y=-\infty$ (recall that the time of flight
to infinity is finite) corresponds to $x = -2+{\cal O}(\beta^{-6/5})$.
Denoting these roots as $y_1(e)\leq y_2(e)\leq y_3(e)$ we find (see
Fig.~(2))
\begin{eqnarray}
y_2(e_c) = y_3(e_c)=-y_1(-e_c) = - y_2(-e_c) = y_0 \ ; \ \ \
y_1(e_c) = - y_3(-e_c)=- 2y_0 \ .            
\end{eqnarray}

The density of states $\rho(\mu_F)=Tr\delta(\mu_F-{\cal H})$ at energy
$\mu_F$ is given in the semiclassical approximations by
\begin{eqnarray}
\rho_{\rm WKB} (\mu_F) = \frac{1}{\pi} \int
\frac{\theta(\mu_F-V_N(x))}{\sqrt{2(\mu_F-V_N(x))}} dx      
\end{eqnarray}
where upon double scaling we find
\begin{eqnarray}
\rho_{\rm WKB} (e) = \frac{\beta^{1/5}}{\pi} \int
\frac{\theta(e-v(y))}{\sqrt{2(e-v(y))}} dy    \ .  
\end{eqnarray}

Since tunnelling effects are unobservable here (as we have remarked at
the end of the previous section) we restrict ourselves to states which
are located semiclassically in the large abyss on the left.  Doing so we
find
\begin{eqnarray}
\rho_{\rm WKB} (e) &=& \frac{2\beta^{1/5}}{\pi} \int^{y_1(e)}_{-\infty}
\frac{dy}{\sqrt{(y_1-y)(y_2-y)(y_3-y)}} = \nonumber \\
&&\frac{4\beta^{1/5}}{\pi} [y_3(e)-y_1(e)]^{\frac{1}{2}}
K\left(\frac{y_3-y_2}{y_3-y_1}\right)                      
\end{eqnarray}
where $K(m)$ is the complete elliptic integral of the first
kind\cite{MHB}.

Note the interesting fact that the contribution to $\rho_{\rm WKB}$ from
the smaller well on the right (i.e., from the integration region
$y_2\leq y\leq y_3$) is exactly the same as our result in Eq.~(20) which
measures the density of states in the region $y\leq y_1(e)$.  This is
clearly a manifestation of supersymmetry at the planar level\cite{MP}.
This implies immediately that the times of flight in both wells are
equal.

As $e\rightarrow e_c+$ ~ we reach the pure gravity point.  Using
Eq.~(17) we find
\begin{eqnarray}
\rho_{\rm WKB}(e_c)=
\frac{4\beta^{1/5}}{\pi} (3y_0)^{\frac{1}{2}} \cdot K(0) = 2\beta^{1/5}
(6t)^{-1/4} = 2[6(1-\frac{N}{\beta})]^{1/4}              
\end{eqnarray}
in accordance with \cite{MP,JF1,JF2}.  $\rho_{\rm WKB}(e)$ diverges as
$\beta^{1/5}$ due to the fact that $e_c$ is the $N$-th level in the
potential $V_N(x)$ where $N\sim\beta\rightarrow\infty$.

Increasing $e$ towards the quadratic maximum $v(-y_0)=-e_c$ (Fig.~(2))
we reach the $c=1$ gravity phase , where
$\frac{y_3-y_2}{y_3-y_1}\rightarrow 1-$.  Using
$K(m) \rightarrow \frac{1}{2}\ln\frac{16}{1-m}+{\cal O}(1)$ as $m
\rightarrow 1-$ \cite{MHB} we find
\begin{eqnarray}
\rho_{\rm WKB}(e\rightarrow\mid e_c\mid-)\simeq\frac{\beta^{1/5}}{\pi}
(3y_0)^{-1/2}\left[\ln \frac{\mid e_c\mid}{\mid e_c\mid - e} + {\cal
O}(1)\right] \ ,                       
\end{eqnarray}
indicating we are in the $c=1$ phase. Using Eq.~(9) the latter result
may be written in the form
\begin{eqnarray}
\rho_{\rm WKB}(e\rightarrow\mid e_c\mid-)\simeq\frac{1}{2\pi}
\left[-V_N^{\prime\prime}(x_c)\right]^{-1/2}
\ln\left(\frac{1}{2}\frac{V_N(x_c)-V_N(x_0)}{V_N(x_c)-\mu_F}\right) 
\end{eqnarray}
in conformity with \cite{MKAZ,C1}. Note that $m$ is a non trivial function
of $e$, namely $m =\frac{y_3(e)-y_2(e)}{y_3(e)-y_1(e)}$ (Eq.~(20)).
Only as $e \rightarrow \mid e_c\mid-$ ~ (i.e. $m \rightarrow 1-$)
does $1-m$ become proportional to $(\mid e_c\mid-e)^{1/2}$ which leads to
the logarithmic divergence in Eqs.~(22)-(23) signifying the $c=1$ string phase.
For $e$ appreciably below $\mid e_c\mid$ we can use the approximation\cite{MHB}
\begin{eqnarray}
K(m) = A_n(1-m)+B_n(1-m)\ln\frac{1}{1-m}+\epsilon(m) \ ;
\ \ 0 < m < 1    
\end{eqnarray}
where $A_n$ and $B_n$ are polynomials of degree $n$ in $1-m$.
The error is bounded by $\mid\epsilon(m)\mid\leq2\cdot10^{-8}$ already
for $n=4$. We then expand $\ln(1-m)$ in powers of $m$ and the logarithmic
behaviour of $K$ and $\rho$ is suppressed, disappearing completely for $m=0$
($e=e_c$). Reducing $e$ {\em below} $e_c$, i.e. below the
Marinari-Parisi point, Eq.~(16) yields one real turning point $y_1(e)$
and two complex conjugate ones $y_{2,3}(e)$ (where $Re \ y_2 > y_1$).
In this case we find
\begin{eqnarray}
\rho_{\rm WKB}(e) = \frac{4\beta^{1/2}}{\pi\mid y_2-y_1\mid^{1/2}}
K\left(\frac{1}{2}-\frac{1}{2} \Re\left\{\frac{y_2-y_1}{\mid
y_2-y_1\mid}\right\} \right) \ , \ \ e < e_c \ .    
\end{eqnarray}

Thus,Eqs.~(21) and (25) imply continuity of $\rho$ at $e_c$:
\begin{eqnarray}
\rho_{\rm WKB}(e_c-) = \rho_{\rm WKB}(e_c+)      \ .   
\end{eqnarray}

We close this section by estimating the total number of particles in the
gas as $e \rightarrow \mid e_c \mid$, the critical point of the $c=1$
phase.  As $\mu_F \rightarrow V_N(x_c)$
we have only two classical turning points located at
$x_L = - 2 + {\cal O}(\beta^{-6/5})$ and $x_R = 2 + 2y_0\beta^{-2/5} +
{\cal O}(\beta^{-4/5})$~ (Fig.~(1)).  The semiclassical expression for the
total
number of particles yields then
\begin{eqnarray}
{\cal N}_c = \frac{\beta}{\pi} \int^{x_R}_{x_L}
\sqrt{2(V_N(x_c)-V_N(x))} dx = N + \frac{2}{\pi}(3y_0)^{1/2}N^{4/5} +
{\cal O}(N^{3/5})     
\end{eqnarray}
i.e. ${\cal N}_c \simeq N \left\{ 1 + \frac{2}{\pi}
[6(1-\frac{N}{\beta})]^{1/4}\right\}$.  Thus, adding approximately
$\frac{2}{\pi}(6t)^{1/4}N^{4/5}\sim\frac{N}{\sqrt{\alpha^\prime}} << N$
particles to the Marinari-Parisi gas.

The result $({\cal N}_c - N)/N << 1$ is clearly consistent with the fact
that the scaling regions of the two phases coincide.\\

\break

\noindent{\bf IV. \ The Two Point Function on the Sphere}\\

Up to this point we have shown the matching of scaling regions and
demonstrated the flow from the $c=1$ gravity phase to the pure gravity
phase by analysing the semiclassical density of states which is
identical to the puncture two point funciton at zero momentum on
the sphere\cite{C1REV,TWOPOINT2,TWOPOINT3}.  This flow manifests itself
further in correllators at non-zero momentum as the chemical potential
$e$ is varied from $-e_c=v(-y_0)$ to $e_c=v(y_0)$.  Here we analyse the
puncture two point function $\langle T \left(\frac{1}{\cal N} Tr
\Phi^r(\tau)\frac{1}{\cal N}Tr\Phi^s(0)\right)\rangle_{\rm conn.}$ on
the sphere.  To this end we must compute the classical trajectory at a
given chemical potential $\mu_F$\cite{TWOPOINT1,TWOPOINT2,TWOPOINT3}.
The classical equation of motion reduced to quadratures is
\begin{eqnarray}
\frac{dx}{d\tau} = \pm \sqrt{2[\mu_F-V_N(x)]} =
\pm \frac{1}{4}\sqrt{(x-x_L)(x_1-x)(x_2-x)(x_3-x)}   
\end{eqnarray}
where the turning points $x_i(i=L,1,2,3)$ are shown on fig.~(1).  We
recall from the previous sections that $x_L\simeq -2+{\cal
O}(\beta^{-6/5})\simeq a$ and $x_i=2+y_i(e)\beta^{-2/5}$ $(i = 1,2,3)$.

Restricting the solution to the large abyss on the left in fig.~(1)
$(x_L\leq x\leq x_1)$, we find the one subjected to the initial
condition $x(0)=x_1$ $(\dot{x}(0)<0)$ to be \cite{MHB}
\begin{eqnarray}
x(\tau) = x_2 -
\frac{(x_2-x_1)(x_2-x_L)}{(x_2-x_L)-(x_1-x_L)sn^2(u\mid m)} \ .  
\end{eqnarray}
Here $sn$ is a Jacobian elliptic function, whose argument $u$, parameter
$m$ and coparameter $m_1$ are
\begin{eqnarray}
u = \frac{1}{8}\sqrt{(x_3-x_1)(x_2-x_L)} \tau \equiv \Omega\tau    
\end{eqnarray}
\begin{eqnarray}
m = \frac{(x_3-x_2)(x_1-x_L)}{(x_3-x_1)(x_2-x_L)} \ , \ \ m_1 = 1-m =
\frac{(x_3-x_L)(x_2-x_1)}{(x_3-x_1)(x_2-x_L)} \ .                
\end{eqnarray}
$x(\tau)$ is a doubly periodic function, whose periods are half the periods
of $sn(u|m)$, namely,
\begin{eqnarray}
T = \frac{2K(m)}{\Omega} \ , \ \ iT^\prime =
\frac{iK(m_1)}{\Omega} \ .          
\end{eqnarray}

$x(\tau)$ degenerates into simpler forms at the critical points $\mu_F =
V_N(x_c)$ and
$\mu_F = V_N(x_0)$ corresponding to $c=1$ gravity $(m=1, \ x_1 = x_2)$
and pure gravity $(m=0; \ x_2=x_3)$, respectively.

The classical trajectory at the pure gravity point can be found simply
by inserting the corresponding turning points into Eq.~(29).  Using
\cite{MHB} $sn(u|0)=\sin u$, Eqs.~(29) and (30) yield
\begin{eqnarray}
x(\tau) = x_2-2
\frac{(x_2-x_1)(x_2-x_L)}{(2x_2-x_1-x_L)+
(x_1-x_L)\cos \omega \tau}                
\end{eqnarray}
where
\begin{eqnarray}
\omega = \frac{1}{4}\sqrt{(x_2-x_L)(x_2-x_1)} =
2\Omega(m=0) = \rho^{-1}_{\rm WKB}(e_c)          
\end{eqnarray}
in agreement with Eq.~(21)\cite{JF1,JF2}.

In order to find the non trivial classical trajectory at the $c=1$ gravity
phase we solve Eq.~(28) for the appropriate parameters\footnote{This is a
singular point of the family of solutions in Eq.~(29) because the real
period (Eq.~(32)) diverges for $x_2=x_1$.  In this case Eq.~(28) implies
for the initial condition $x(0)=x_1$ that $\ddot{x}(0)=0= -\partial
{V_N(x_1)}/ \partial{x}$ as well as $\dot{x}(0)=0$,hence $x(\tau) = x_1={\rm
const.}$   We therefore solve Eq.~(28) with another initial condition,
namely, $x(0)=x_L, \ \dot{x}(0) > 0$ } and find (see fig.~(1))
\begin{eqnarray}
x(\tau) = x_1- \frac{2\frac{(x_1-x_L)(x_3-x_1)}{x_3-x_L} }
{\cosh \left( \frac{\tau}{4}\sqrt{(x_1-x_L)
(x_3-x_1)}\right) + \frac{x_3+x_L-2x_1}{x_3-x_L} } \ ,
x(0)=x_L \ .                     
\end{eqnarray}

As we approach the $c=1$ gravity critical point,
$\mu_F\rightarrow V_N(x_c) -$ \ , Eq.~(31) implies $m\rightarrow1-, \
m_1\rightarrow0+$.

Thus, using the expansions\cite{MHB} $sn(u|m)\simeq\tanh u +
\frac{1}{4}m_1 (\sinh u \cosh u - u)\sec h^2u+{\cal O}(m^2_1)$ and
$cn(u|m)\simeq\sec hu -\frac{1}{4}m_1(\sinh u \cosh u - u)\tanh u \cdot
\sec hu + {\cal O}(m^2_1)$ as $m_1\rightarrow 0 +$ \ in Eq.~(29) we
find
\begin{eqnarray}
x(\tau) \simeq x_1 + \frac{x_3-x_1}{x_3-x_L}
(x_1-x_L)\frac{m_1}{2} [1-\cosh 2u] +{\cal O}(m^2_1) \ . 
\end{eqnarray}
Note that this approximation is valid almost along the whole period
$|\tau|\leq T/2$ (i.e.$|u|\leq K(m))$ which is a better situation than the
one cosidereed in\cite{TWOPOINT3}\footnote{We note from Eq.~(28) that
$\ddot{x}(0)<0$
while $\ddot{x}(T/2)>0$.  As is clear form Eq.~(2) and from fig.~(1),
$V^\prime_N(x)$ vanishes at $x\simeq(-1)$.It can be shown by inverting
Eq.(29)that near the c=1 gravity point the particle spends most of the period
near the right critical turning point.Therefore the particle reaches the
point $x=-1$ for the first time (given the initial condition of Eq.~(29)) at
$\tau\simeq T/2$. The approximate solution in Eq.~(36), on the other hand
implies
$\ddot{x}(\tau)<0$ for any $\tau$.}.

The Euclidean two point function
\begin{eqnarray}
G(\tau) = \langle \left(\frac{1}{\cal N} Tr\Phi^r(\tau)\frac{1}{\cal N}
Tr\Phi^s(0)\right)\rangle_{\rm conn.}               
\end{eqnarray}
where $r$ and $s$ are positive
integers\cite{TWOPOINT1,TWOPOINT2,TWOPOINT3}is given by
\begin{eqnarray}
G^{(r,s)}(\tau) = \sum^\infty_{n=1} nf^{(r)}_n f^{(s)}_n
e^{-n\omega|\tau|}                                   
\end{eqnarray}
to leading order in $1/\beta$, i.e. on the sphere.\\

Here $\omega = 2\pi/T$ and
\begin{eqnarray}
f^{(r)}_n=\frac{1}{T} \int^{T/2}_{-T/2} d\tau \ x^r(\tau)\cos(n\omega\tau) \ ,
\
\ n \geq 1                                        
\end{eqnarray}
are the cosine Fourier components of $x^r(\tau)$.

The monentum space correlator is therefore
\begin{eqnarray}
G^{(r,s)}(p) = \sum^\infty_{n=1} \frac{n^2\omega f_n^{(r)}f_n^{(s)}}
{p^2+n^2\omega^2}                                   
\end{eqnarray}
(where $p$ is conjugate to $\tau$).

At this point let us note for later convenience that Eqs.~(30) and (31)
expressed in terms of the scaled coordinate $y$ become
\begin{eqnarray}
u\simeq\frac{1}{4}(y_3-y_1)^{1/2}\beta^{-1/5} \tau
\left(1+{\cal O}(\beta^{-2/5})\right) =
\left(\frac{y_3-y_1}{3y_0}\right)^{1/2}
\frac{\tau}{\sqrt{4\alpha^\prime}}                 
\end{eqnarray}
\begin{eqnarray}
m \simeq
\frac{y_3-y_2}{y_3-y_1}
\left(1+{\cal O}(\beta^{-2/5})\right) \ , \ \
m_1 =
\frac{y_2-y_1}{y_3-y_1}
\left(1+{\cal O}(\beta^{-2/5})\right)                 
\end{eqnarray}
in accordance with Eq.~(20).

Let us consider the $c=1$ gravity phase first.  The approximation to the
classical trajectory (which we take as valid virtually along the whole
period $|\tau|\leq T/2)$ in Eq.~(36)
coincides with the one found in \cite{TWOPOINT3} up to parameter
redefinitions.  Rewriting Eq.~(36) in terms of double scaled coordinates
(Eqs.~(41) and (42)) we find
\begin{eqnarray}
x(\tau)\simeq x_1+\sqrt{(2\alpha^\prime\mu)}\left\{1-\cosh\left[ \left(\frac
{y_3-y_1}
{3y_0}\right)^{1/2}\frac{\tau}{\sqrt{\alpha^\prime}}\right]\right\}
\end{eqnarray}
where we have used the fact that
\begin{eqnarray}
y_2-y_1 \simeq 4\sqrt{\frac{2}{3}\frac{|e_c|
-e}{y_0}} {\rm  ~ as~}  e\rightarrow\mid e_c \mid -
\end{eqnarray}                                                 
from Eqs.~(10)and(16).

Calculating the relevant Fourier components in the validity domain of
Eqs.~(36) and (43) \footnote{Note\cite{TWOPOINT3}that only the $\tau$
dependent part of Eq.(43) is relevant for the scaling part of G} we find
\begin{eqnarray}
f^{(r)}_n = \frac{(-1)^n\omega}{\pi}
\left(\frac{\alpha^\prime\mu}{2}\right)^{r/2} \sum^r_{k=0}
\left(\matrix{r \cr k}\right)\frac{\left(\frac{r-2k}{\sqrt{\alpha^\prime}}
\xi\right)
\sinh\left(\frac{r-2k}{\sqrt{\alpha^\prime}}\xi T\right) }
{\left(\frac{r-2k}{\sqrt{\alpha^\prime}}\xi\right)^2+n^2\omega^2}  
\end{eqnarray}
where $\xi = \left(\frac{y_3-y_1}{3y_0}\right)^{1/2}$ and $n$ is any
positive integer\footnote{The result in Eq.~(45) is obtained by using Eq.~(43)
in Eq.~(39).  The integration is carried along the validity domain of the
former which is virtually the whole period $|\tau|\leq T/2$.} ~\footnote{in
the $c=1$ phase we may safely
substitute $\xi\simeq 1$.  Otherwise we may replace $\alpha^\prime
\rightarrow \alpha^\prime/\xi^2$ in Eqs.~(46)-(48) below}.

Following \cite{TWOPOINT2} we may sum Eq.~(40) over all integers as
\begin{eqnarray}
G^{(r,s)}(P) = \frac{1}{2i} \oint dz \frac{\omega z^2}{p^2+\omega^2z^2}
f^{(r)}_zf^{(s)}_z\cot(\pi z)        
\end{eqnarray}
where the integration contour wraps around the real axis.  Rotating the
contour to wrap around the two halves of the imaginary axis (neglecting
an infinite but $\omega$ independent contribution from the cotangent),
we may evaluate $G(p)$ by summing the appropriate residues.

The simple poles at $z=\pm ip$ lead to techyonic contributions, namely
\cite{TWOPOINT1,TWOPOINT2,TWOPOINT3}
\begin{eqnarray}
G^{(r,s)}_{\rm tach.}(p) = F_{r,s}(p)\cdotp\coth(pT/2) \ .      
\end{eqnarray}

$F_{r,s}(p)$ is the usual form factor which depends explicitly on the
operators involved in Eq.~(37), contains (a finite number of) simple and
double poles at real integral values of $\sqrt{\alpha^\prime}\cdot p$
and is regular at $p=0$.  The other part of Eq.~(47) is the universal
tachyonic contribution.
The latter is proportional to the usual expression
$p\cdot(|e_c|-e)^{\sqrt{\alpha^\prime}p}$~\footnote{Using
Eqs.~(24),(32),(42) and (44) we find
$T\simeq\sqrt{\alpha^\prime}\mid\ln(|e_c|-e)\mid$ as we expect for the
$c=1$ gravity phase.  Furthermore, approximating
$\coth(pT/2)\simeq1+2e^{-pT} \ (pT>>1)$ we obtain the expression in the
text.}

The other poles in Eq.~(46) are either simple or double ones for generic
values of $p$~~\footnote{By generic values of $p$ we mean those that are
non integral multiples of $\frac{1}{\sqrt{\alpha^\prime}}$.
Otherwise, we may get third order
poles in Eq.~(46) as well if $\mid p\sqrt{\alpha^\prime}\mid\leq
\min\{r,s\}$.  Such poles are ignored in \cite{TWOPOINT3} despite the
fact that they may lead to terms proportional to $\mid\ln\mu\mid^2$.}.
They are associated with the special states of $c=1$ gravity
\cite{TWOPOINT3,SPECIAL}.  Only the double poles yield contributions to
$G(p)$ which are enhanced by the singular factor $T\sim|\ln\mu|$ coming
from derivatives of the cotangent in Eq.~(46).

These contributions are found to be
\begin{eqnarray}
\frac{\sqrt{\alpha^\prime}}{4\pi}
\left(\frac{\alpha^\prime\mu}{2}\right)^{\frac{r+s}{2}}
\mid \ln(\sqrt{\alpha^\prime}\mu)\mid \sum^r_{k=0}
\left(\matrix{r \cr k}\right) \left(\matrix{s \cr
\frac{1}{2}(s-r)+k}\right)
\frac{(r-2k)^2}{\alpha^\prime p^2-(r-2k)^2}       
\end{eqnarray}
where it is assumed that $r\leq s$, which agrees with
\cite{TWOPOINT3}.

Eqs.~(43)-(48) establish therefore that the two point correlators of the
matrix model under consideration possess all features of $c=1$ gravity
as $\mu_F\rightarrow V_N(x_c)$ (i.e. $e\rightarrow|e_c|-)$.

This conclusion further supports our results for these correlators at
zero momentum, namely -- the density of states, in the previous section.

We turn now to the pure gravity phase where the matrix model coincides
with the Marinari-Parisi model.  The classical trajectory is given in
Eq.~(33).  We will compute $G^{(1,1)} (\tau)$ for simplicity.  In this case
Eq.~(39) yields
\begin{eqnarray}
f_n \equiv f^{(1)}_n = x_2\delta_n_,_o +2
\frac{(x_2-x_1)(x_2-x_L)}{x_1-x_L} (-1)^n
\frac{e^{-n\alpha}}{\sinh \alpha}              
\end{eqnarray}
where
\begin{eqnarray}
\sinh^2\alpha =
\frac{4(x_2-x_1)(x_2-x_L)}{(x_1-x_L)^2}
=
\frac{3y_0\beta^{-2/5}(1+\frac{1}{4}y_0\beta^{-2/5})}
{(1-\frac{1}{2}y_0\beta^{-2/5})^2 }             
\end{eqnarray}
and $x_i$ are the turning points corresponding to pure gravity (Eq.~(17)
and Fig.~(1)).

Inserting the last two equations into Eq.~(38) we obtain the desired
correlator in closed form:
\begin{eqnarray}
G^{(1,1)}(\tau) = \left(1-\frac{y_0}{2} \beta^{-2/5}\right)^2
\left[\frac{\sinh\alpha}{\sinh (\alpha + \frac{\omega\tau}{2})}\right]^2
\ ,               
\end{eqnarray}
where $\omega$ is given in Eq.~(34).  We see immediately that
$G^{(1,1)}(0)\simeq
1-y_0\beta^{-2/5}=1-(\frac{2t}{3})^{1/2}\beta^{-2/5}$ gives the correct
pure gravity two puncture correlator on the sphere\cite{DSL} as we
expect from the formalism of stochastic stabilisation on general
grounds\cite{GH}, and in accordance with \cite{MP,KM,AMB,AMB1,JF1,JF2}.

The scaled chemical potential at the pure gravity point (measured from
the quadratic maximum $v(-y_0))$ is $\mu = 2|e_c|\sim t^{3/2}$.  The
(renormalised) cosmological constant $t$ is therefore related to $\mu$
at this point via $t\sim\mu^{2/3}$, in contrast to the $c=1$ gravity
relation $t\sim\mu|\ln\mu|$.

There are clearly no tachyonic contributions to the fourier transform of
$G^{(1,1)}(\tau)$ since the classical period is finite and does not
depend logarithmically on $|e_c|-e$.
Thus,$G^{(1,1)}(\tau)$ really describes pure
gravity -- i.e., the case of no embedding dimension.

Up to this stage we have calculated $G^{(r,s)}(\tau)$ very close to the
two critical points, and found that it behaves as expected at each
phase.  For intermediate values of $\mu_F$ in the range
$V_N(x_0)<\mu_F<V_N(x_c)$, calculation of $G^{(r,s)}(\tau)$
becomes cumbersome and
will not be pursued here further.

The latter calculation is essential to determination of the nature
of the phase transition under investigation.In particular, it would be
very interesting to compare it with the BKT transition in the
compactified model,especially due to the fact that the critical
compactification radius diverges here as
$\sqrt{{\alpha^\prime}}\simeq\beta^{1/5}$.Finally note that in principle
multi-puncture correlators at zero momentum can be calculated exactly and
to any order in the genus expansion by using the diagonal resolvent of the
one particle hamiltonian h \cite{KM,C1REV,TWOPOINT2,TWOPOINT3}generalising
our analysis in section (III).
\pagebreak

\noindent{\bf V. \ Conclusion}\\

By generalising the stochastically stabilised matrix model of Marinari
and Parisi we have obtained a hybridised situation in which the scaling
regions of $c=1$ gravity and pure gravity coincide.  This occurs due to
the extremly large  Regge slope $\alpha^\prime$ we obtain in the $c=1$
gravity phase, which scales in a {\em specific} way in $\beta$. As a result
we may vary the chemical potential over a finite domain and flow
continuously from the $c=1$ gravity phase to the pure gravity one,
all in the {\em singlet} sector of the matrix model.
This has been demonstrated by investigating the changes in two puncture
correlators at zero momentum (i.e. --density of states) and at non-zero
momenta as the chemical potential has been varied. Since the flow we have
been discussing here does not involve non-singlet sectors it seems to be in
contrast to the BKT transition occuring in the compactified matrix model.
The fact that the critical compactification radius of the latter becomes
infinite for the $\alpha^\prime$ scale involved here may,though,shed some
light on the connection between these two cases.We have left open the
question wheter the singlet sector partition function of our model is dual
under inversion of the compactification radius.

Since the flow from the $c=1$ phase into the pure gravity phase which
implies dissipation of the string embedding dimensions occurs due to the
fact that $\alpha^\prime \sim \beta^{2/5} \rightarrow \infty$ it is
interesting to check whether it is a two dimensional stringy manifestation
of the high energy behaviour of critical string theory\cite{GROMENDE}
in which the unbroken extremely large symmetry of string theory is beleived
to resurect\cite{GROWITT} and space-time looses its ordinary meaning.
Indeed,considerations made in \cite{GROMENDE}seem to be sensitive only to
parameters of worldsheets that dominate the string path-integral
expressions for scattering amplitudes and are basicly innert to the quantum
numbers of the scattered string states and to the embedding dimension.
We should expect for a proper manifestation of theses effects in our case
eventhough most of the string modes have been degenerated into discrete states
in two dimensions except for the massless tachyon.In particular-we should
expect attenuation of,say,tachyon scattering amplitudes as we reach the pure
gravity phase,since the tachyon mode does not exist there.Since these
amplitudes are automatically high energy amplitudes on the energy scale set
by $\alpha^\prime$ in our model this attenuation should be consistent with
\cite{GROMENDE}. Note,however,that all our discussion has been based on
a specific matrix model which has been constructed in order to give a well
defined definition of pure gravity.Nevertheless,if the flow discussed here is
somehow related to its counterpart in the compactified model despite the
fact that it involves only singlet states, we have a rather sound basis to
its universality. At this point we note,however,that high energy behaviour
of two dimensional string theory has been investigated in\cite{MOORE,RONEN}
whose conclusions contradict those of\cite{GROMENDE},namely,the former find
that high energy amplitudes are polynomial in momenta and do not decay
exponentially as the latter predict.It seems,however,that our specific
matrix potential for which the scaling regions of the two gravitational
phases overlap, may evade the conclusions of\cite{MOORE,RONEN} for the
reasons stated above.\\

\noindent{\bf Acknowledgements}\\

Interesting discussion with D. Bar-Moshe, M.S. Marinov and M. Moshe are
kindly acknowledged.  We are espcially indebeted to M. Moshe for his
support and interest along all stages of this work.

We also thank Y. Ben-Horin for help in preparation of the figures.
\pagebreak

\noindent{\bf Figure Captions}\\

\noindent{\em Fig. (1)}: The potential $V_N(x)$ for $\frac{N}{\beta} =
0.5$.\\

\noindent{\em Fig. (2)}: The scaled potential $v(y)$ for $t=1.5$.
\pagebreak

\end{document}